\documentclass{emulateapj}
\shorttitle{Brightest cluster galaxies}
\shortauthors{Ruszkowski \& Springel}

\begin{document}
\title{The role of dry mergers for the formation and evolution of brightest
cluster galaxies}              
\author{M. Ruszkowski\altaffilmark{1,2} and V. Springel\altaffilmark{2}}

\altaffiltext{1}{Department of Astronomy, The University of Michigan, 500 Church Street, Ann Arbor, MI 48109, USA;\\
E-mail: mateuszr@umich.edu}
\altaffiltext{2}{Max Planck Institute for Astrophysics,
Karl-Schwarzschild-Str. 1, 85741 Garching, Germany;\\
E-mail: volker@map-garching.mpg.de}

\begin{abstract}
  Using a resimulation technique, we perform high-resolution
  cosmological simulations of dry mergers in a massive
  (10$^{15}M_{\odot}$) galaxy cluster identified in the {\it
    Millennium Run}. Our initial conditions include well resolved
  compound galaxy models consisting of dark matter halos and stellar
  bulges that are used to replace the most massive cluster progenitor
  halos at redshift $z=3$, allowing us to follow the subsequent dry
  merger processes that build up the cluster galaxies in a self-consistent
  cosmological setting.  By construction, our galaxy models obey the
  stellar mass-size relation initially. Also, we study both galaxy
  models with adiabatically contracted and uncompressed halos.  We
  demonstrate that the brightest cluster galaxy (BCG) evolves away
  from the Kormendy relation as defined by the smaller mass galaxies
  (i.e., the relation {\em bends}).  This is accompanied by a
  significantly faster dark matter mass growth within the half-light
  radius of the BCG compared to the increase in the stellar mass
  inside the same radius. As a result of the comparatively large
  number of mergers the BCG experiences, its total mass-to-light ratio
  becomes significantly higher than in typical elliptical galaxies.
  We also show that the mixing processes between dark matter and stars
  lead to a small but numerically robust tilt in the fundamental
  plane and that the BCG lies on the tilted plane. 
Our model is consistent with the observed steepening of 
the logarithmic mass-to-light gradient as a function of the stellar mass.
As we have not included effects from gas dynamics or star
  formation, these trends are exclusively due to $N$-body and
  stellar-dynamical effects. Surprisingly, we find 
  only tentative weak
  distortion in the Faber-Jackson relation that depends on the aperture size, unlike expected based on
  studies of isolated merger simulations. This may be due to
  differences in the distribution of galaxy orbits, which is given in
  our approach directly by the cosmological context while it has to be
  assumed in isolated merger simulations, and the fact that the BCG is
  located deep in the cluster potential well. Another uncertainty in both
  approaches lies in the definition of the spatial extent of the BCG.
\end{abstract}


\section{Introduction}

\subsection{What makes BCGs special ?}

The brightest cluster galaxies (BCGs) are special. They are the most massive
and luminous galaxies in the Universe. They are typically located in the very
centers of clusters of galaxies which indicates that their formation is
closely linked to that of the clusters themselves.  Their formation history is
therefore distinct from typical elliptical galaxies (e.g., Lin \& Mohr 2004,
Brough et al. 2005, De Lucia \& Blaizot 2006 and references therein).  Recent
analysis by Best et al. (2007) shows that BCGs are also more likely to host
active galactic nuclei than other galaxies of the same stellar mass. This
indicates that these objects play a pivotal role in quenching cooling flows
and star formation in clusters.  

BCGs have been recently shown to lie off the standard scaling relations of
early-type galaxies (Lauer et al. 2007, von der Linden et al. 2007, Bernardi
et al. 2007). In particular, they show excess luminosity (or stellar mass)
above the prediction of the standard Faber-Jackson relation at high galaxy
masses. This hints at the interesting possibility that their black hole masses
may be larger than estimated from the $M-\sigma$ relation (Tremaine et
al. 2002).

\subsection{Formation of BCGs in the cosmological context}

Early theoretical studies suggested that BCGs are formed as a result of star
formation in {\it cooling flows} (Fabian 1994).  However, recent X-ray
observations performed with {\it Chandra} and {\it XMM Newton} show that
cooling rates in nearby clusters are too low to explain the masses of BCGs.
Tidal stripping and dynamical friction acting on the cluster galaxies have
been suggested as an alternative BCG formation mechanism (Oemler 1976,
Richstone 1976) but this mechanism turns out to be too slow due to the low
cross section for galaxy-galaxy interactions in virialized cluster centers
where relative velocities of galaxies are large (Ostriker 1980). Nevertheless,
this {\it cannibalism} may be responsible for the formation of diffuse stellar
envelopes extending up to hundreds of kiloparsecs (the `cD' galaxies).

However, in a hierarchical structure formation theory, groups of
galaxies will form before the formation and virialization of a
cluster. Such groups have sufficiently small velocity dispersions that
a significant number of galaxy-galaxy mergers occurs.  This process
should preferentially lead to the formation of elliptical galaxies in
high galaxy density environments.  Stellar population synthesis models
demonstrate that the bulk of star formation in most massive galaxies
took place prior to $z\sim 2$ (Thomas et al. 2005, Treu et al. 2005,
Jimenez et al. 2006).  Semi-analytical models also show that stars
that make up most of the BCG mass are formed very early on (80\% by
$z\sim 3$; De Lucia et al. 2006, De Lucia \& Blaizot 2007).  Only
after most stars have been formed does the final galaxy assembly take
place (the {\it downsizing} phenomenon; 
Bundy et al. 2005, Pannella et al. 2006, Calura et al. 2008, but see
the revised monolithic collapse model of Chiosi \& Carraro (2002) and 
the results of Cimatti et al. 2006 who, based on the same COMBO-17 and DEEP2 data as Bell et al. 2004,
argue for ``top-down'' assembly of early-type galaxies, i.e., that 
downsizing concept may have to be 
extended to the mass assembly itself as the build-up of the most massive galaxies precedes 
that of the less massive ones.).  
Therefore, the final galaxy mergers in cluster formation are expected to be predominantly
dissipationless (or {\it dry}, i.e.~not involving large amounts of
gas; Khochfar \& Burkert 2003). This picture is also consistent with the analysis based on
COMBO-17 results of Bell et al.~(2004) who show that the stellar mass
on the red sequence increased passively by a factor of $\sim 2$ since
$z\sim1$. We note that recent theoretical results suggest that the amount of gas 
associated with major mergers in the early Universe was much larger then at the present epoch
(Khochfar \& Silk 2006). Such {\it wet} mergers led to massive starbursts early on. However,
at lower redshifts the amount of gas was much lower and the dominant mechanism for 
the galaxy growth in mass and size were dry mergers (e.g., Hopkins et al. 2008). 
This is evidenced indirectly by the strong size evolution of the most
massive galaxies since $z=2$ (Trujillo et al. 2007) while preserving the 
mass-size relation found by Trujillo et al. (2004).

Despite these findings, we note that the detailed picture of the giant 
elliptical assembly is further constrained
by photometric and chemical observables. They suggest that 
it is difficult to explain giant elliptical by a pure sequence of multiple minor dry mergers or 
via major dry mergers (i.e., some admixture of wet mergers may be required
to properly model chemical observables; Pipino \& Matteucci 2008).
However, the scatter in $\sigma $--[Mg/Fe] relation 
is large enough to permit one to three major dry mergers 
during the galactic lifetime (Bell et al. 2006).
While the majority of BCGs from
the sample studied by Bildfell et al. (2008)  have bluer colors with
increasing radius, about 25\% possess bluer cores. The existence of such cores indicates 
that some cold gas must have been replenished in that subsample of 
BCGs at late epochs via minor wet mergers.
Interestingly, Bildfell et al. (2008) find that 
the Kormendy relation of the BCGs is steeper than that of the local ellipticals.
Kaviraj et al. (2008) argue that 
bright early-types potentially form 10-15 per cent of their stellar mass since z = 1
and, according to Pipino et al. (2008),
the recent star formation in the blue core BCGs typically has an age less than 
0.5 Gyrs and contributes mass fractions of less than a percent.
von der Linden et al. (2007) show that BCGs exhibit
higher [$\alpha$/Fe] values than same mass ellipticals and claim that
this indicated that star formation may have occurred over a shorter
timescale in  BCGs. This then suggests a possibility that 
the [$\alpha$/Fe] radial gradients in BCGs could be consistent with being 
comparable to those in same mass ellipticals (Brough et al. 2005) 
rather then being unable to survive repeated dry mergers.

We also point out that the AGN feedback plays an important role in 
explaining the early and  rapid formation of most massive ellipticals
by shutting off star formation. The effect of AGNs may be felt more strongly
in more massive galaxies that have been heated by virial shocks. This made their gas 
more dilute and more vulnerable to AGN feedback
leading to passive evolution and `red-and-dead' massive 
spheroids (Dekel \& Birnboim 2006). Recent high-resolution {\it FLASH} 
code simulations of the interaction of relativistic jets with the ISM
confirm that the star formation rate will be inhibited although it 
can increase on short time-scales of the order of $10^{5}-10^{6}$ yr 
Antonuccio-Delogu \& Silk 2008).

As mentioned above, recent analysis of the {\it SDSS} and {\it Hubble} data
demonstrates that BCGs form a separate class of objects that depart from the
usual Faber-Jackson relation. This effect has been seen in numerical
simulations of dry mergers performed by Boylan-Kolchin et al. (2006) who
studied collisions of isolated galaxy pairs.  They demonstrated that the
initial orbital parameters play an important role in shaping the Faber-Jackson
relation.  In their model, galaxies that collide on more radial orbits, such
as those present during the cluster assembly, lead to more puffed up merger
remnants. Such final merger products are characterized by lower internal
velocity dispersions than those resulting from lower ellipticity collisions.
This mechanism is consistent with the fact that most massive galaxies 
experienced at least one merger since $z<1.5$ (Conselice et al. 2007).
As noted by Lauer et al. (2007), the increase in the ratio of the velocity
dispersion of galaxies moving in the cluster potential to the internal
velocity dispersion of individual cluster galaxies may also contribute to the
puffing up effect. This is because some of the orbital energy of the colliding
galaxies will be transferred to the internal energy of the merger
remnant. Another effect that may influence the slope of the Faber-Jackson
effect is tidal heating of galaxies moving in the cluster potential.  It is
due to all of the above complexities that one has to resort to cosmological
simulations involving both the dark matter and stellar components to capture
all relevant processes leading to the formation of BCGs. It is the purpose of
this paper to address some of the above issues. We will deliberately use
purely collisionless simulations, allowing us to obtain reliable $N$-body
results in a fully cosmological context without having to delve into the
complexities of gas dynamics and star formation. However, we note that the
approach we outline here can be used also in future hydrodynamic simulations
of the cluster formation process.

The paper is organized as follows. In Section 2, we outline the numerical
approach used to simulate the cluster and its galaxies.  Section 3 discusses
our results (scaling relations, tilt in the fundamental plane, the evolution
of the mass-to-light ratio, and merger rates). Then in Section 4, we discuss
the numerical convergence and the dependence of our results on the initial
conditions and the galaxy models. Finally, Section 5 summarizes our findings.

\section{Methods}

Our approach to simulate the evolution of the BCG is similar in spirit to that
of Dubinski (1998), who replaced 100 dark halos at $z=2$ in a proto-cluster
region with galaxy models constructed in isolation, and then continued the
simulation to the present epoch. However our method differs in several
important details, and for definiteness, we list its primary steps in the
following:

\begin{itemize}

\item We identify a massive $10^{15}M_{\odot}$ cluster at $z=0$ in the
  {\it Millennium Run} (Springel et al. 2005a) and track its particles back to the
  initial conditions.

\item We determine the Lagrangian region of the cluster material in the
  original unperturbed initial conditions and produce a new refined
  unperturbed particle load at $z=127$ which is adequate for a {\em zoom
    simulation} onto the selected target halo at higher resolution.

\item We generate the actual initial conditions by perturbing the particle
  load obtained in the previous point with the Millennium displacement field,
  augmented with additional small-scale waves that can now be added due to the
  improved mass resolution of $\simeq 10^{9}h^{-1}M_{\odot}/F_{\rm
    zoom}^{3}$. We consider $F_{\rm zoom}=2$ and $F_{\rm zoom}=4$ in this
  paper, corresponding to approximately 8 and 64 times the original
  resolution.  Further away from the cluster, the resolution is progressively
  downgraded in mass, and the gravitational softening lengths are enlarged,
  ensuring however that the tidal forces acting on the high-resolution region
  are represented accurately at all times.

\item We evolve these initial conditions to $z=3.0$.

\item We identify the fifty most massive dark matter halos at this redshift
  (almost all of them end up in the final cluster) and replace them with
  galaxy models at either the same mass resolution as in the high-resolution
  region, or with the resolution enhanced by another factor of eight. The
  models consist of both dark matter and stellar particles (see below). In
  every run the dark matter and star particles of galaxy models have the same
  mass in order to minimize differential two-body heating effects between the
  two components.  The maximum mass and force resolutions for each of our
  simulations are listed in Section 3.

\item We evolve the cluster from $z=3.0$ to the present epoch.

\end{itemize}

We note that this approach is also similar in spirit to that recently employed by
Rudick, Mihos \& McBride (2006) who focused on the problem of the
intracluster light and studied significantly less massive ($\sim 10^{14}M_{\odot}$) clusters.
We only consider spheroidal galaxy models in this paper that include a
stellar bulge and a dark matter halo.  The bulge size obeys the size-stellar
mass relation of Shen et al. (2003):
\begin{equation}
R_{e}=4.16h^{-1}\left(\frac{M_{\rm
star}}{10^{11}h^{-1}M_{\odot}}\right)^{0.56}{\rm kpc},
\end{equation}

\noindent
where $R_{e}$ is the half-light radius, and $M_{\rm star}$ is the
stellar mass of a given galaxy. There is observational evidence
(Trujillo et al. 2004, McIntosh et al. 2005) that this relation
remains unchanged up to $z\sim 1$ as long as the most massive ellipticals
are excluded. Note that more recent observations suggest that the size-luminosity
relation evolves with redshift for high-mass ellipticals (Zirm et al. 2007,
van Dokkum et al. 2008) but at these mass scales this trend is naturally 
explained  by our simulations as explained in section 3). For the sake of simplicity we here
assume that the relation also holds at our redshift ($z=3$) for
constructing the initial galaxy models (see Section 3 for 
more discussion of this assumption).  We describe the distribution
of mass in stars $m_{*}(r)$ in the bulge by a Hernquist profile, i.e.,
\begin{equation}
\rho_{*}(r) = \frac{M_{*}}{2\pi}\frac{a}{r(r+a)^{3}},
\end{equation}
where $a$ is a characteristic radius, $M_{*}$ is the total mass of the stellar
component in a given galaxy and $r$ is the distance from its center. We
truncate the Hernquist profile at $r=10\, a$ in our numerical realizations in
order to avoid the occurrence of a very small number of very distant star
particles. For this choice of parameters, the half-light radius is
$R_{e}=1.339\,a$.

We assume that 90\% of the original dark matter mass of each halo excised from
the zoomed Millennium simulation consists of dark matter and the rest
contributes to the stellar component.  The dark matter halos are modeled with
the NFW profile (Navarro, Frenk \& White 1996) truncated at $r_{200}$, with a
concentration parameter $c_{\rm vir}$ given by a modified version of the
Bullock et al. (2001) prescription,
\begin{equation}
c_{\rm vir}(\mu,a)=K\frac{a}{a_{c}},
\label{EqnConcentration}
\end{equation}
where $K$ is a constant, $a=1/(1+z)$, $a_{\rm c}$ is the epoch of collapse of
a given halo and $\mu=M_{\rm vir}(a)/M_{*}(a)$ is a ratio of a halo mass and
the typical halo mass at the same epoch $a$.  The typical mass at an epoch $a$
is defined as $\sigma[M_{*}(a)]=1.686/D(a)$, where $D(a)$ is the linear growth
factor and $\sigma(M)$ is the rms density fluctuation on a comoving scale
containing the mass $M$. The collapse epoch is defined by
$M_{*}(a_{c})=FM_{\rm vir}$, where $F$ is a free parameter.  In the
computation of $c(\mu,a)_{\rm vir}$, as well as in all of our simulation runs
(carried out with the {\small GADGET-3} code; Springel 2005c), we assume the
cosmological parameters of the Millennium Run, given by $\Omega_{\rm m}=0.25$,
$\Omega_{\Lambda}=0.75$, a scale invariant slope of the power spectrum of
primordial fluctuations ($n=1.0$), a fluctuation normalization
$\sigma_{8}=0.9$, and a Hubble constant of $H_0 = 100\,h\,{\rm km\,
  s^{-1}\,Mpc^{-1}} = 73\,{\rm km\, s^{-1}\,Mpc^{-1}}$. The constants $K=3.4$
and $F=0.01$ have been chosen following 
Macci{\`o} et al. (2006).  We note that
recently Gao et al.~(2008) have provided a yet more accurate parametrization
of the concentration dependence on mass and redshift than given by equation
(\ref{EqnConcentration}), but our results are insensitive to the size of this
revision.

Once the density structure of our galaxy models has been specified,
velocity distributions of the bulge and dark matter particles are
obtained by solving the Jeans equations following the method of
Springel et al. (2005b). For the sake of simplicity, we assume that
the dark matter halos have zero net angular momentum.

Where indicated, the dark matter distribution has been modified by
including adiabatic contraction following the classic Blumenthal et
al. (1986) approximation. Hydrodynamical simulations by Gnedin et
al.~(2004) suggest that this approximation overpredicts the amount of
contraction at small radii. We note however that these differences
between hydrodynamical simulations and the contraction model are smaller
than the differences between the uncompressed and contracted halos,
and that an overestimate of the contraction effect allows us to {\em
  bracket} reality between the two models we consider.

Nevertheless, we want to stress that our toy galaxy models are too
simplistic to represent real galaxies very accurately.  However, the
advantage of our approach is that it gives us better control over the
parameters of the galaxy population under study and allows us to
cleanly separate the pure $N$-body effects from the complicated
physics of gas dynamics, radiative cooling, star formation, etc. Our
method should still be sufficiently accurate to investigate the
evolution of the main trends in galaxy scaling relation, especially in
the $K$-band.  While the luminosity of the early type galaxies on the
red sequence does evolve with redshift, the passive stellar evolution
does not change the $K$-band luminosity much and hence should not
dramatically affect our results. This approach differs from, e.g., that of
Conroy et al. (2006) and Berrier et al. (2008) who combine dark
matter simulations with recipes for assigning light to dark matter
halos and study the assembly of a large number of clusters, or from
that of Murante et al.  (2007) who incorporate the gasdynamical
effect and star formation at the expense of lower dynamical range.\\
\indent
In our analysis, the BCG was identified using first the
friends-of-friends algorithm to find the cluster halo and its galaxies,
and then by applying the SUBFIND group finding algorithms to look for
gravitationally bound structures. We did not set any strict isophotal
limit when defining galaxies, instead we let SUBFIND identify the bound
stellar and dark matter part. This sets a certain isophotal limit for
the BCG.  The corresponding isophotal limit for non-BCGs may be slightly
different but that should not affect their half-light radii
significantly as they do not possess extended envelopes.  Moreover,
unless the change in the isophotal limits is huge, it is not likely to
substantially affect the total luminosities and half-light radii of the
galaxies in our sample.

\section{Results}

All results presented in this section correspond to our default case
where the initial galaxy models were adiabatically contracted. For
details of the numerical parameters for all runs see Section 4.
Figure 1 shows the evolution of the relation between the half-light
radius $R_{e}$ and the stellar mass within that radius. This relation
is similar to the Kormendy relation. The initial ($z=3$) and final
($z=0$) models are represented by the red and black points,
respectively. Intermediate redshifts are marked by open circles with
colors as given in the inset. The figure demonstrates
that, while smaller mass galaxies show relatively little evolution,
the BCG clearly gains a substantial amount of mass and increases in
size. Interestingly, it clearly evolves off the extrapolation of the
initial radius-mass relation. We stress that, in the current scenario,
this effect is entirely due to $N$-body process and is not caused by
gas-dynamical processes, radiative cooling, star formation,
etc. This effect has been reported in the observational work of Lauer et al. (2007).
They showed that the half-light radii of BCGs lie above the log-log 
relation between the half-light radius and stellar luminosity 
defined by smaller mass galaxies. They also attribute this effect to a "progressive
change in the character of `dry mergers' at higher galaxy masses".
This has also been reported even more recently by 
Bildfell et al. 2008 who state that ``the Kormendy relation of our 
BCGs is steeper than that of the local 
ellipticals, suggesting differences in the assembly history of these 
types of systems." We also note that the trend for the 
size of the most massive galaxies to evolve with redshift (Zirm et al. 2007, van Dokkum et al. 2008) 
is consistent with the evolution of our simulated relation.
Interestingly, our results also imply the steepening of 
the logarithmic mass-to-light gradient as a function of the stellar mass.
The  logarithmic $M/L$ gradient (Napolitano et al. 2005)
evaluated using $r_{\rm in} = R_{\rm e}$ and   $r_{\rm out} = R_{\rm
vir}$ is approximately  $\nabla_{l}\gamma \approx (R_{\rm e}/R_{\rm
vir})[(M_{\rm DM}/M_{s})_{\rm out}- (M_{\rm DM}/M_{s})_{\rm in}]\propto  R_{\rm
e}/R_{\rm vir}$ because $R_{\rm e}\ll R_{\rm vir}$, $(M_{\rm
DM}/M_{s})_{\rm out}\gg (M_{\rm DM}/M_{s})_{\rm in}$  and  $(M_{\rm
DM}/M_{s})_{\rm out}$ is a constant ($M_{\rm DM}$ and $M_{s}$ denote the 
dark matter and stellar masses enclosed within a given radius, respectively). 
Since $R_{\rm e}\propto
M_{s}^{\alpha}$, we have $\nabla_{l}\gamma \propto M_{s}^{\alpha
-1/3}$. Now, for $\alpha =0.56$ the relation is flat while for higher
stellar masses $\alpha$ increases and so does the logarithmic $M/L$
gradient. This is in agreement with the observational findings of
Napolitano et al. (2005). 

Figure 2 shows the evolution of the stellar (violet) and dark matter
(red) components within the half-light radius of the BCG as a function
of redshift. The values on the vertical axis are normalized to the
dark matter mass within the half-light radius at $z=3$.  It is evident
from this figure that the bend in the $R_{e}-M$ relation is
accompanied by a significant gain in mass of the BCG both in terms of
the stellar and dark matter components. Interestingly, the ratio of
the dark matter-to-stellar mass increases with time significantly. In
other words, the regions of the BCG within the half-light radius
become disproportionally more massive with time for a given stellar luminosity.
We note that the stellar mass accretion rates of BCGs (a factor of $\sim 2.5$
since $z\sim 2$) obtained by Romeo et al. (2008) are consistent with our findings.

The change in the dark matter-to-stellar mass ratio is likely caused by the
large number of mergers that the BCG experiences over time, especially when
compared to other galaxies. In Figure 3 we show a histogram of the number of
mergers experienced by the BCG, the second largest and the third largest
galaxy in the cluster. A merger is here defined as an event where over 50\% of
the stellar mass from one galaxy ends up in another galaxy. The vertical axis is
for the cumulative number of mergers from the initial redshift $z_{\rm ini}=3$
down to a given redshift $z<z_{\rm ini}$ (shown on the horizontal axis).  As
is clearly seen, the number of mergers declines rapidly with galaxy mass, and
most of the mergers happening in the cluster are between the BCG and smaller
galaxies. 

We note that the results for the most massive galaxies are quite
insensitive to the exact definition of the $n$-th most massive galaxy,
which can be defined as either (i) $n$-th largest at a  given redshift
or (ii) $n$-th largest at $z=0$. Strictly speaking, we adopt the latter
definition and backtrace the history of the particles belonging to
such-defined $n$-th largest galaxies in the computational volume. The
fact that the results for the most massive objects do not depend on the
choice of the definition shows that the most massive galaxies preserve their
identities and it is mostly smaller objects that accrete onto
them. The Figure 3 indeed confirms that the BCG experiences
overwhelmingly more mergers than smaller galaxies. This fact is the
key reason for the
change in the mass ratio between dark matter and the stellar component
within the BCG's half-light radius. The number of mergers experienced 
by the most massive galaxy is zero between $z=2$ and $z=3$ and the smaller the galaxy 
mass the lower the redshift at which it experiences a merger (if any).
Therefore, the results are not sensitive to the 
exact insertion redshift for the galaxy models.

We note that the boost in the dark matter to stellar mass ratio could be
contaminated by mergers of small starless halos (which were not replaced with
galaxy models) with a bigger halo. However, the dynamical friction time is a
strongly decreasing function of the mass of the object that is merging with a
bigger one. Therefore, the contribution to the growth in the BCG mass that
comes from the small halos that we do not populate with stars is expected to
be weak. This fact is not inconsistent with the above observation that the
most massive objects do not merge with the BCG. It simply means that among the
objects within a certain mean distance from each other, the most massive ones
are the first ones to merge. In other words, the biggest galaxies in our
simulations have simply been prevented from merging thus far by their large
distance.

Curiously, the stellar mass-$\sigma$ dispersion shown in Figure 4 does
not show strong
%any sign
bend similar to that presented in Figure 1.
Figure 4 shows the evolution of the stellar mass within the half-light
radius and the velocity dispersion $\sigma$ which, in our model, is
essentially an equivalent of the Faber-Jackson relation.  The initial
slope of this relation is $\sim 3.62$.  The meaning of the symbols is
the same as in Figure 1.  The stellar velocity dispersions are
measured within half-light radii, are three-dimensional (i.e., all
three components of velocity are used to compute them) and are
intensity-weighted (i.e., in our case, weighted by the stellar mass
integrated along the line-of-sight). At the final redshift the BCG
lies on the relation defined by the extrapolation of the initial
galaxy models to more massive objects.\\ \indent The right panel shows
the flux-weighted $\sigma$ as a function of stellar mass for three
different apertures: half-light radius (red points), 1/8 or the
half-light radius (green points) and 1/16 of that radius (blue
points). There is a systematic trend for the BCG to move to smaller
velocity dispersions as the size of the aperture is decreased. At the
same time, low-$\sigma$ objects only increase their scatter around the
unaltered mean values. There is also a tentative weak hint of a {\it bend}
in this relation but this conclusion depends on the precise
measurement of lowest $\sigma$ galaxies that show scatter and the
magnitude of the bend may be slightly affected by the 
precise definition of the BCG boundary that separates it from the
intracluster light (ICL).

We also note that the dependence of sigma 
on luminosity is slightly more subtle. In the case of size-luminosity relation, 
for a given 
binding energy of the merging pair, the satellites on high angular orbits 
need to lose more angular momentum to merge and then end up being more compact. Those
on lower angular momentum orbits are then more puffed up. On the other hand,
the velocity dispersion close to the galactic center is strongly influenced by the 
stellar bulge and declines with radius up to a point where the cluster potential 
starts to dominate
leading to the increase of the dispersion with the distance from the BCG center.
This means that the cluster potential
may try to slow down the increase of the slope with luminosity of the 
Faber-Jackson relation.
In any case, fitting the extended wings of
the BCG light distribution, beyond of the extent of this central galaxy 
as identified in our on-the-fly algorithm, could perhaps increase the 
bend in this relation 
(and the Kormendy relation) slightly and allow one to study the
evolution of the ICL but such effects should not dramatically alter our 
conclusions. We defer such detailed analysis to a forthcoming
publication.

Liu et al. (2008) analyzed a sample of BCGs and ellipticals
and found that the slope of both the size-luminosity and the Faber-Jackson
relation changes with the galaxy luminosity.
However, they also analyzed BCGs and bright ellipticalls falling into the same 
magnitude range. 
Curiously, they found that, while "the power-law indices for the
size-luminosity relations for BCGs are steeper then those for bright
non-BCG elliptical galaxies, the power-law indices for the
Faber-Jackson relations are also steeper, but only at $\sim 1\sigma$
levels, and so are not statistically significant" (see their Figure
14). This trend appears to be consistent with our 
simulations. The departure of the BCG from the original log-log 
size-luminosity relation increases with time and the slope of the 
size-luminosity for the BCG also seems to rise.  
This clearly suggests that an ensemble of simulated clusters would
show a different size-luminosity slope for BCG and non-BCG galaxies in
the same luminosity range, but a reliable determination of the size of
the effect would require a much larger simulation set.

Finally, Figure 5 shows the ``fundamental plane'' corresponding to the initial
models at $z=3$ (red points) and the galaxies at the final redshift $z=0$
(violet points). The ``brightness'' parameter is $I_{e}=m_{\rm stars}/(\pi
R_{e}^{2})$, where $m_{\rm stars}$ is the stellar mass within the half-light
radius. The absolute values on the axes are not important here as we are only
interested in the evolution of the slope of this relation. The initial slope
of the relation is $\sim 0.83$ (by definition, the virial relation corresponds
to 1.0).  Indeed, there appears to be a small but numerically robust tilt in
the fundamental plane. The final slope of the relation is $\sim 0.72$.
The BCG seems to lie on the tilted plane but, strictly
speaking, this conclusion depends on the robustness of the values measured for
the least massive galaxies (see below for a discussion of numerical
convergence). Nevertheless, the trend for the most massive object(s) to move
away from the original relation is real.  We stress again that this effect is
entirely due to stellar dynamical/$N$-body effects.

\section{Numerical convergence and model robustness}

We performed several test runs in order to assess the numerical convergence of
our results and their sensitivity to the initial galaxy models. For testing
the basic numerical convergence we in general used uncontracted galaxy models;
these have less particles inside the half-light radii and therefore yield more
conservative convergence tests.

Figure 6 shows one of the results of these tests. It presents the relation
between the half-light radius and the stellar mass within that radius.
Initial models are denoted by open red circles. All other points are for
the galaxies at the final redshift $z=0$. Different sets of points grouped by
color correspond to different resolutions. In this figure, we show the results
for four different runs:

\begin{itemize}

\item run A: maximum mass resolution at the baseline level of the zoom
  simulation equal to $1.25\times 10^{8}h^{-1}M_{\odot}$, combined with a
  maximum force resolution of $2\,h^{-1}$kpc.

\item run B: baseline mass resolution unchanged, but the mass resolution of
  the galaxy models enhanced further by a factor of eight.

\item run C: same as above, but the gravitational softening length for the
  particles in the galaxy models was decreased by a factor of two.

\item run D: baseline mass resolution enhanced further by a factor of eight to
  $1.50\times 10^{7}h^{-1}M_{\odot}$, mass resolution of the galaxy models
  equal to the baseline mass resolution, and a gravitational force resolution
  equal to $1\,h^{-1}$kpc everywhere in the high-resolution region.

\end{itemize}

The runs A to D show progressively improving agreement as the resolution of
the simulations is increased. This is especially evident for the smallest
objects.  The comparison of the runs C and D shows that the effects of
two-body relaxation do not play an important role even in the lower resolution
run C.  The maximum resolution run exhibits a close agreement of the galaxy
properties at $z=0$ with those assumed at the initial redshift. As in the case
of Figure 1, this figure shows a clear bend in the Kormendy relation but this
time for the galaxy models that were not
adiabatically compressed.

In Figure 7, we show how the mass ratios evolve as a function of stellar
mass. The color coding is the same as in Figure 1. The left panel is for the
uncompressed initial models and the right one for the adiabatically contracted
ones. As expected, the adiabatically contracted models have higher dark matter
to stellar mass ratios.  Nevertheless, the conclusions remain the same. The
obvious trend for the BCG to evolve in a dramatically different fashion than
the rest of the population in the cluster is present. The conclusion regarding
all other plots discussed above also hold for the models that were not
adiabatically compressed.

The mass of the cluster of galaxies considered here ($10^{15}M_{\odot}$) is
much greater then the total mass in the entire computational volume ($\sim
4\times 10^{13}M_{\odot}$) considered in the cosmological simulations of the
fundamental plane and scaling relations presented by O{\~n}orbe et
al. (2006). Their simulations include the effects of gas cooling and star
formation. However, the mass resolution in our case is much better than the
resolution of $2\times 64^{3}$ particles these authors considered. It is not
clear if their results are numerically fully converged at $z=0$ as their
convergence tests were performed at $z=1$ due to the very high CPU
requirements.  O{\~n}orbe et al. (2006) do not report a departure of the most
massive object from the scaling relations but they do see a tilt in the
fundamental plane. In their model, the tilt is created during an early violent
phase of rapid cooling and star formation. This tilted fundamental plane is
then preserved in the dissipationless mergers that follow.

We acknowledge that we can afford the very high numerical resolution and
dynamical range we achieve in our cluster only at the expense of not including
gas dynamical effects but, as a result, this allows us to address a very clean
and well-defined problem and to isolate the purely dynamical effects from the
complicated gas dynamical processes. Also, in our approach we can ensure by
construction that the galaxies at the initial high redshift obey a prescribed
mass-size relation, which further helps to make the problem we examine clean
and well defined, and to arrive at robust conclusions.

\section{Conclusions and Discussion}

We have performed high dynamical range cosmological $N$-body simulation of the
formation of a massive $10^{15}M_{\odot}$ cluster of galaxies. We included
both the stellar and dark matter particle components in order to investigate
the (differential) evolution of these two components and the relationship
between them. We also performed numerical convergence tests and checked the
robustness of our approach with respect to details of the initial galaxy
models by considering both regular and adiabatically compressed galaxy models.

In our approach, the Kormendy relation (defined here as the relation between
the half-light radius and the mass in the stellar component) evolves with
redshift entirely due to stellar dynamical/$N$-body processes. More
specifically, the BCG clearly departs from the extrapolation of the initial
relation to higher mass objects and the final relation appears to be bent at
the massive end. The evolution of the BCG is accompanied by a systematic and
strong increase in the dark matter-to-stellar mass ratio which is brought
about by a significantly larger number of mergers that the BCG experiences
compared with other galaxies.  In contrast, smaller galaxies show rather
little evolution in this proxy of the mass-to-light ratio.  We speculate that
the above trend also implies that the slope of the $R_{e}-L$ relation for the
BCGs is steeper than for the smaller early-type galaxies. This possibility
would have to be explored in future simulations involving more clusters of
varying mass.  Interestingly, a gentle trend for this quantity to increase
with the galaxy mass has been reported by Gallazzi et al. (2006) in early-type
galaxies in the SDSS survey. The bend in the Kormendy relation has also been
interpreted by Lauer et al. (2007) as being due to an increase in this mass
ratio. We note that such variations may also have important implications for
the lensing properties of the BCGs.

Surprisingly, the Faber-Jackson relation 
only shows a tentative weak bending
with redshift in our simulations that depends on the aperture size.  However, isolated merger
experiments performed mostly for parabolic merger orbits (Boylan-Kolchin et
al. 2006) suggest that such an effect could be present.  We speculate that the
key factors responsible for this difference are: (1) the mergers in our model
all take place in the central cluster potential and the BCG lies very close to
it and (2) our initial conditions for the mergers are self-consistently taken
into account as we evolve all galaxies in the cosmological context starting
from high redshift prior to the cluster assembly. We also note that the
distribution of the initial orbital parameters of the merging galaxies is a
very sensitive function of eccentricity. More precisely, the distribution
possesses a narrow spike at zero eccentricity (Benson 2005), and small
departures from the idealized case of parabolic orbits are then likely in a
realistic scenario, and (3) the precise definition of where the BCG envelope
ends and the ICL component begins may somewhat modify the measured strength of any bend
in these relations.

\section*{Acknowledgments}

MR acknowledges support from the {\it Chandra} theory grant TM8-9011X.  MR
thanks Anja von der Linden, Monica Valluri, Marta Volonteri, Mike
Boylan-Kolchin, Massimo Dotti, Ole M{\"o}ller, Eric Bell and Simon White for many
stimulating discussions and Klaus Dolag for technical help in the early stages
of this project.  The runs were performed on the OPA cluster at the
Rechenzentrum Garching of the Max-Planck-Society and on the {\it Columbia}
supercomputer at NASA NAS Ames center. It is MR's pleasure to thank the staff
of NAS, and especially Johnny Chang and Art Lazanoff, for their highly
professional help. We thank the referee for a very constructive report that 
greatly helped to improve the paper.

%\bibliography{agnpaper} 

\clearpage

\begin{figure*}
\includegraphics[width=0.90\textwidth]{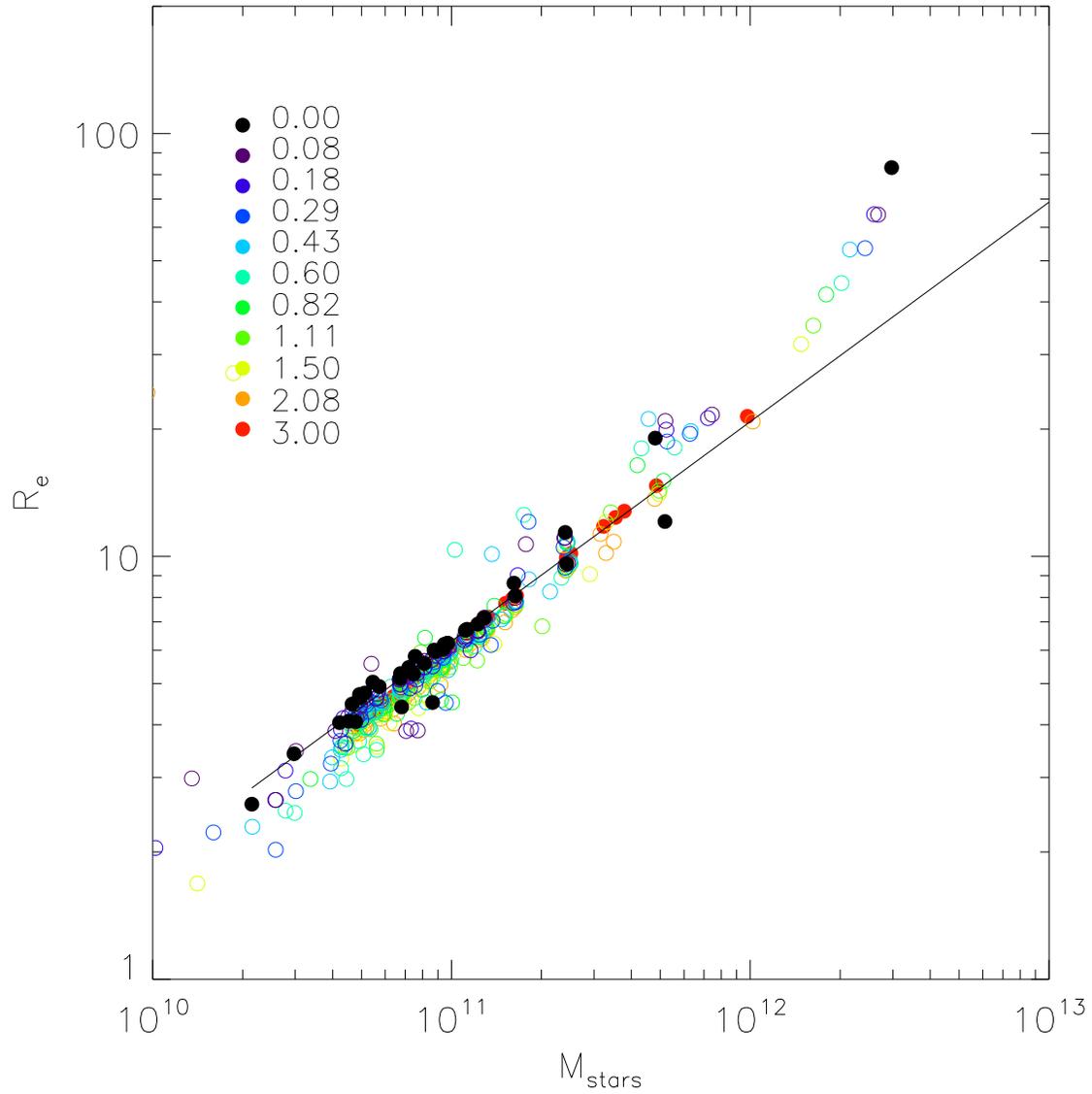}
\caption{Half-light radius vs. stellar mass relation. Filled red
circles are for the initial models at $z=3$ and the black ones are for
$z=0$. Intermediate redshift data is shown as open circles and
distributed in redshift according to the color scale given in the
upper left corner.} 
\end{figure*}

\begin{figure}
\includegraphics[width=0.450\textwidth]{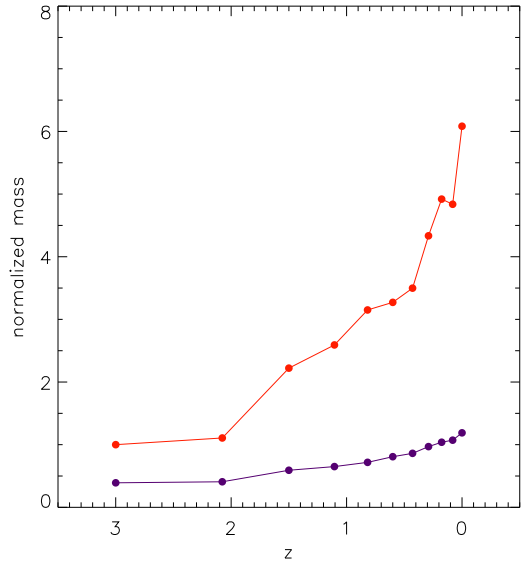}
\caption{Mass in the dark matter (red) and stellar (violet) components
for the BCG as a function of redshift. The values are normalized to
the dark matter mass within the half-light radius at $z=3$.} 
\end{figure}

\begin{figure}
\includegraphics[width=0.450\textwidth,angle=0]{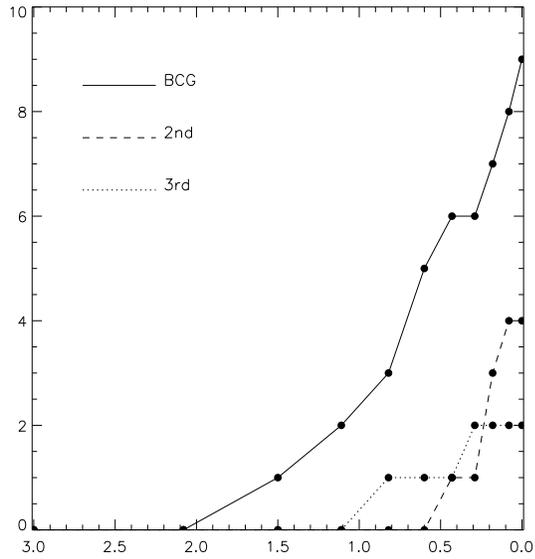}
\caption{Cumulative number of mergers experienced by the three most massive
galaxies in the cluster as a function of redshift.} 
\end{figure}

\begin{figure*} 
\includegraphics[width=0.50\textwidth]{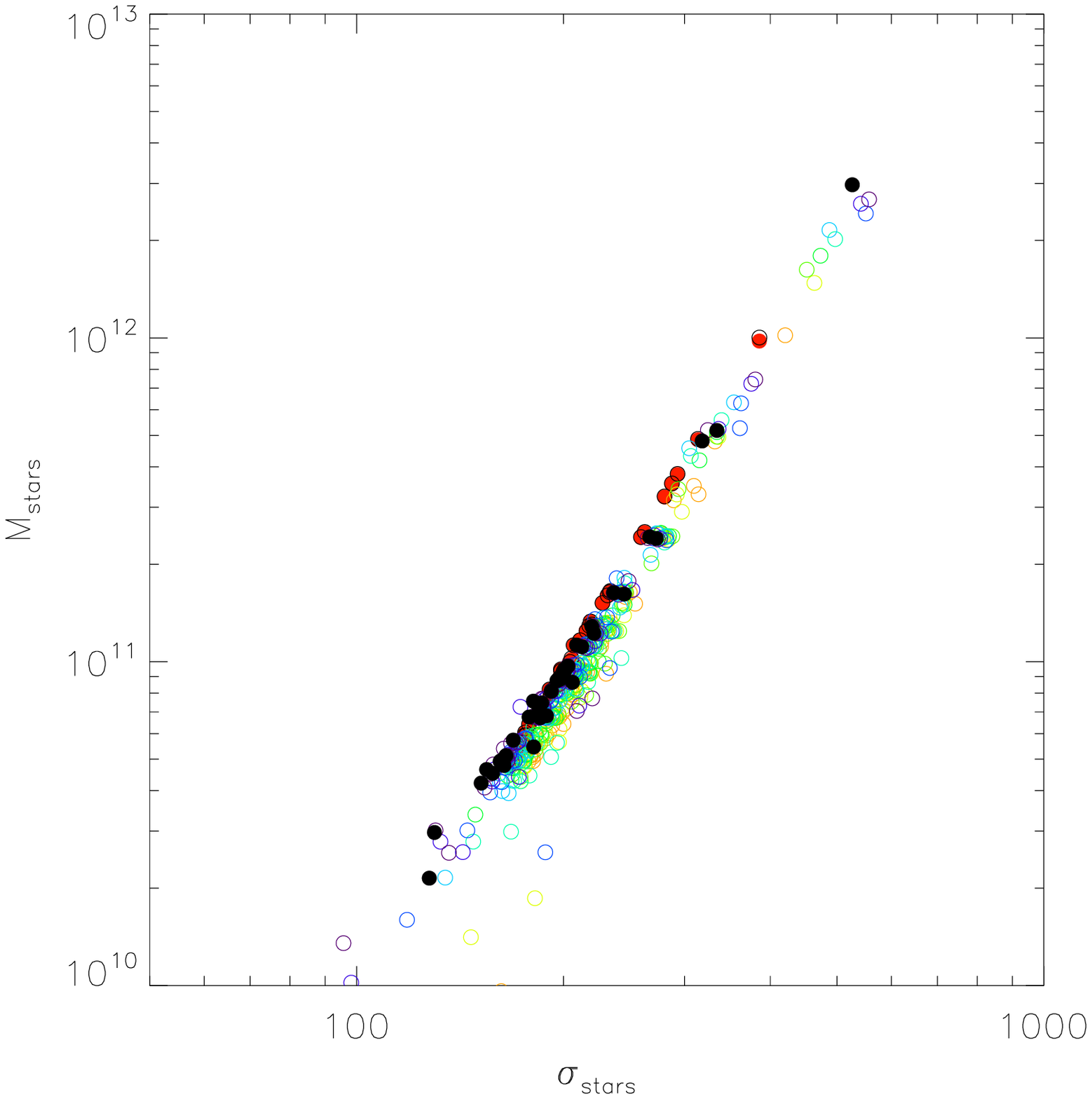}
\includegraphics[width=0.50\textwidth]{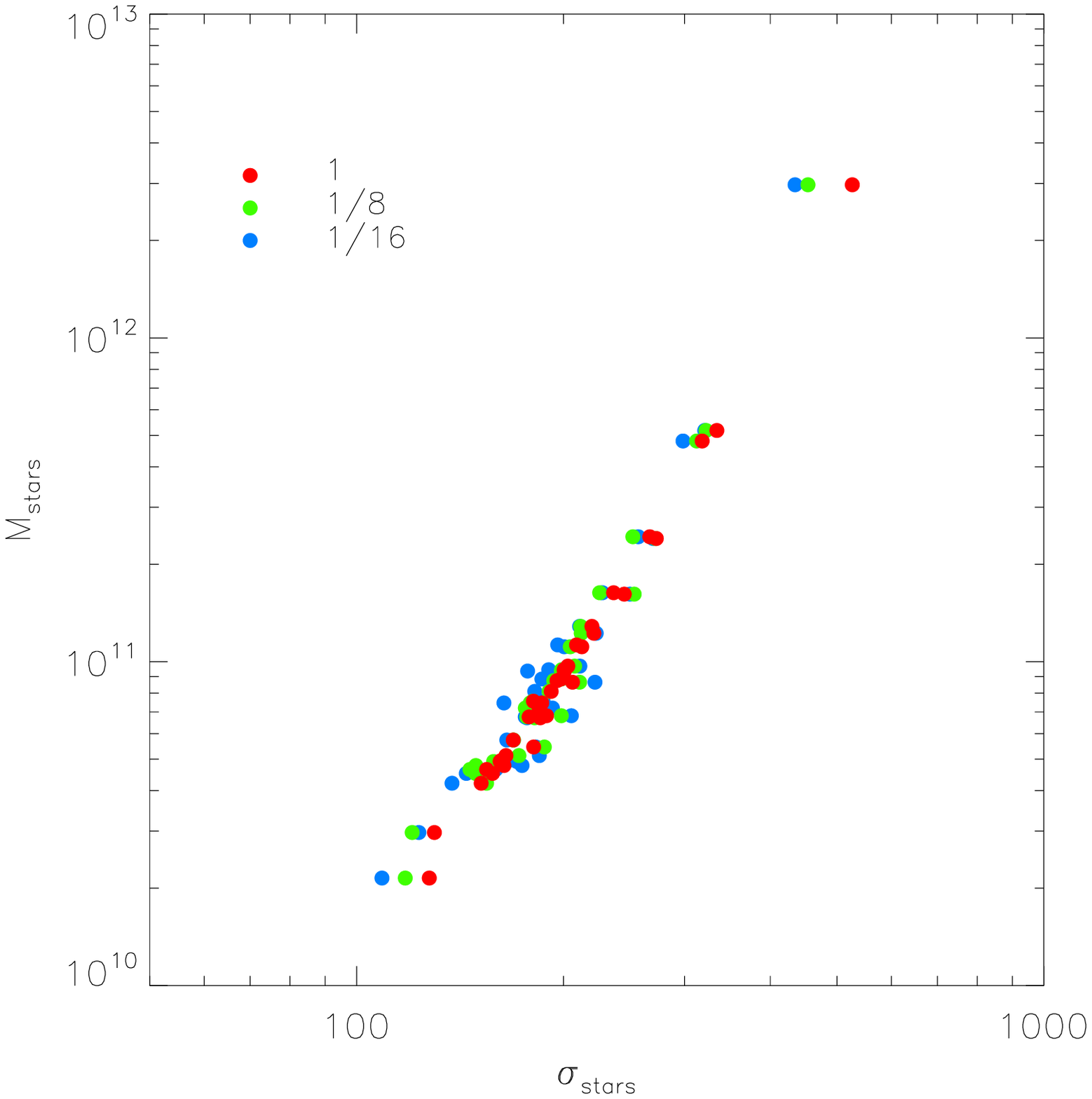}
\caption{Left panel: stellar mass as a function of the flux-weighted
line-of-sight stellar velocity dispersion within the half-light
radii. The meaning of symbols is the same as in Figure 1. Right panel:
flux-weighted line-of-sight stellar velocity dispersion computed for
three different apertures (half-light radius, 1/8 and 1/16 of that
radius) as a function of the stellar mass at $z=0$.}
\end{figure*}

\begin{figure}
\includegraphics[width=0.450\textwidth]{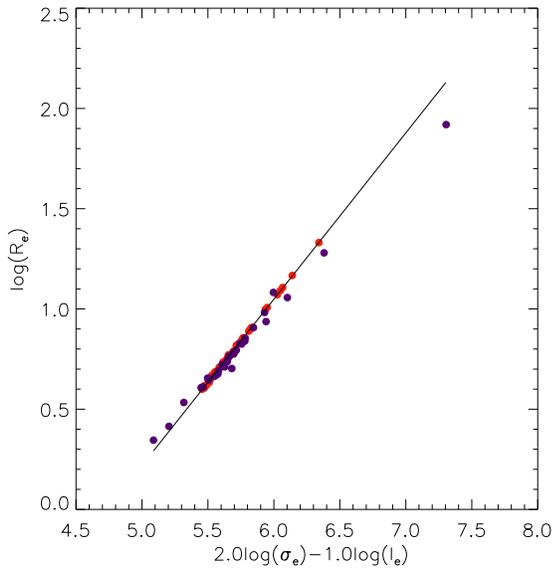}
\caption{Fundamental ``plane'' for the galaxy model at $z=3$ (red) and
$z=0$ (violet). $R_{e}$ and $\sigma_{e}$ are the half-light radius and
the velocity dispersion defined as in Figure 4. The intensity is defined as
$I_{e}=m_{\rm star}/(\pi R_{e}^{2})$.}
\end{figure}

\begin{figure*}
\includegraphics[width=0.90\textwidth]{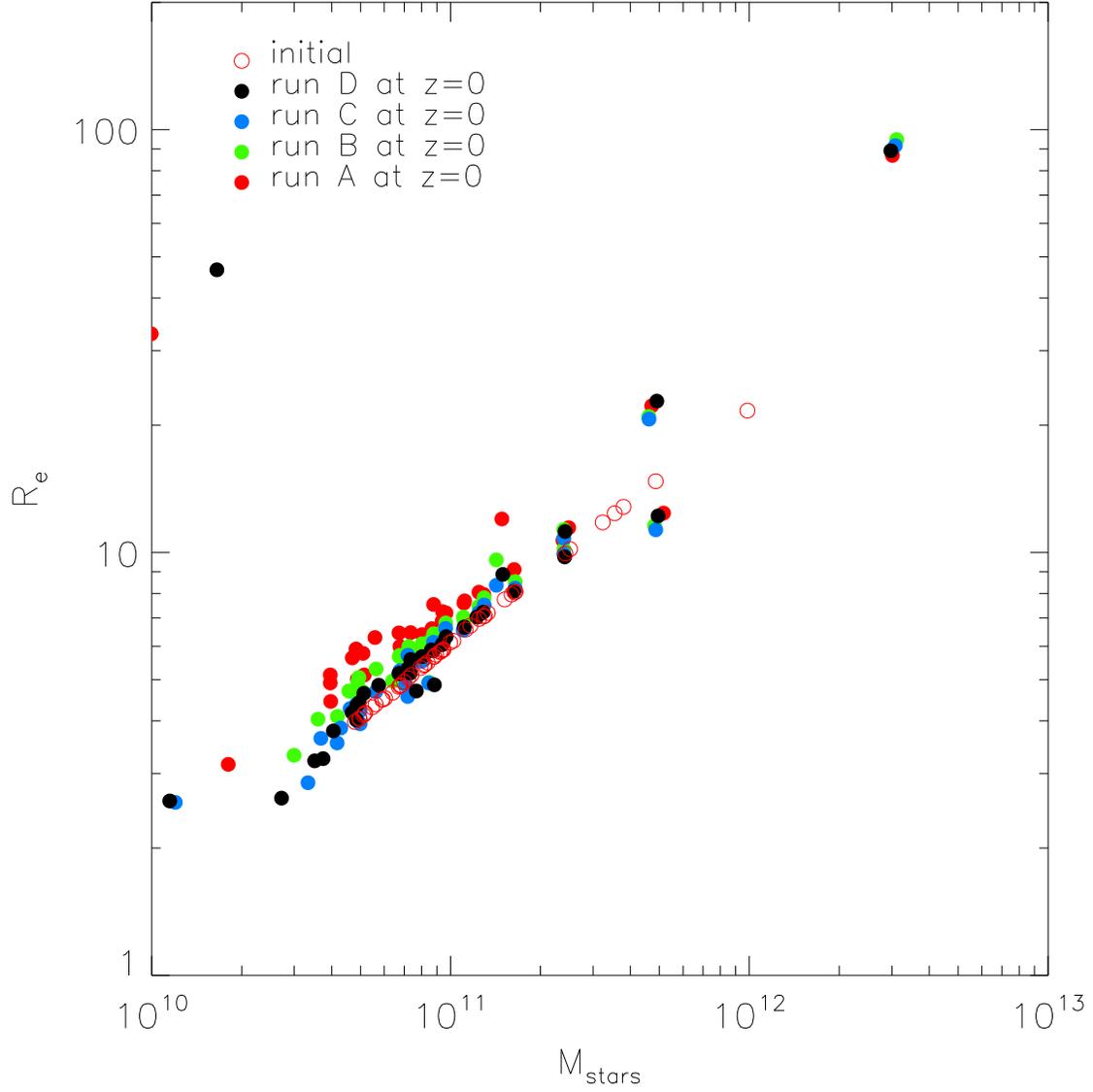}
\caption{Convergence study for the relation between half-light radius and
  stellar mass, $R_{\rm e}-M_\star$, for four different runs of increasing
  numerical resolution (from run A to Run D; see text for details). All filled
  circles correspond to the galaxies at $z=0$. Open red circles are for the
  initial galaxy models at $z=3$. All runs were carried out for uncontracted
  galaxy models.}
\end{figure*}

\begin{figure*}
\includegraphics[width=0.50\textwidth]{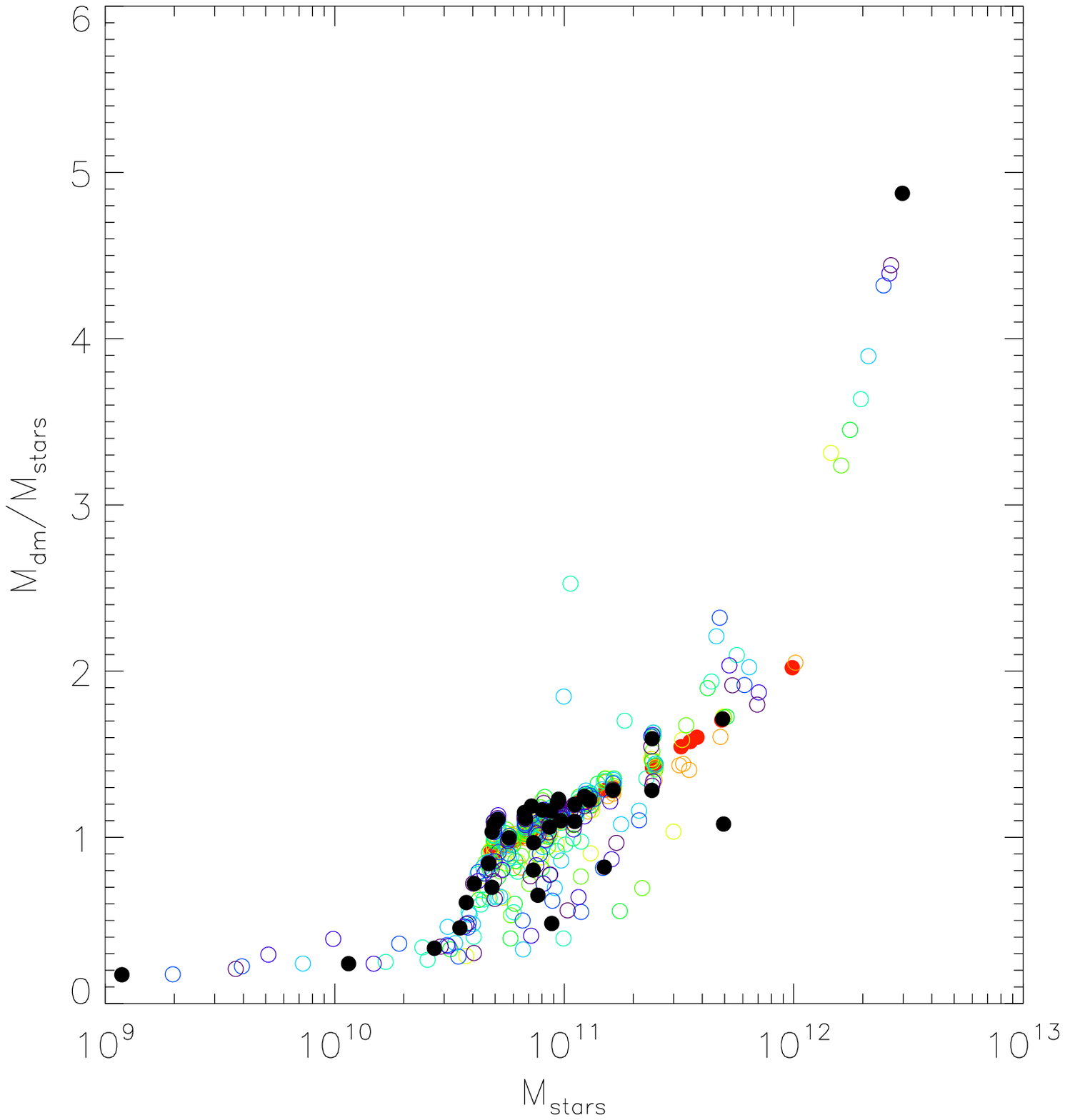}
\includegraphics[width=0.50\textwidth]{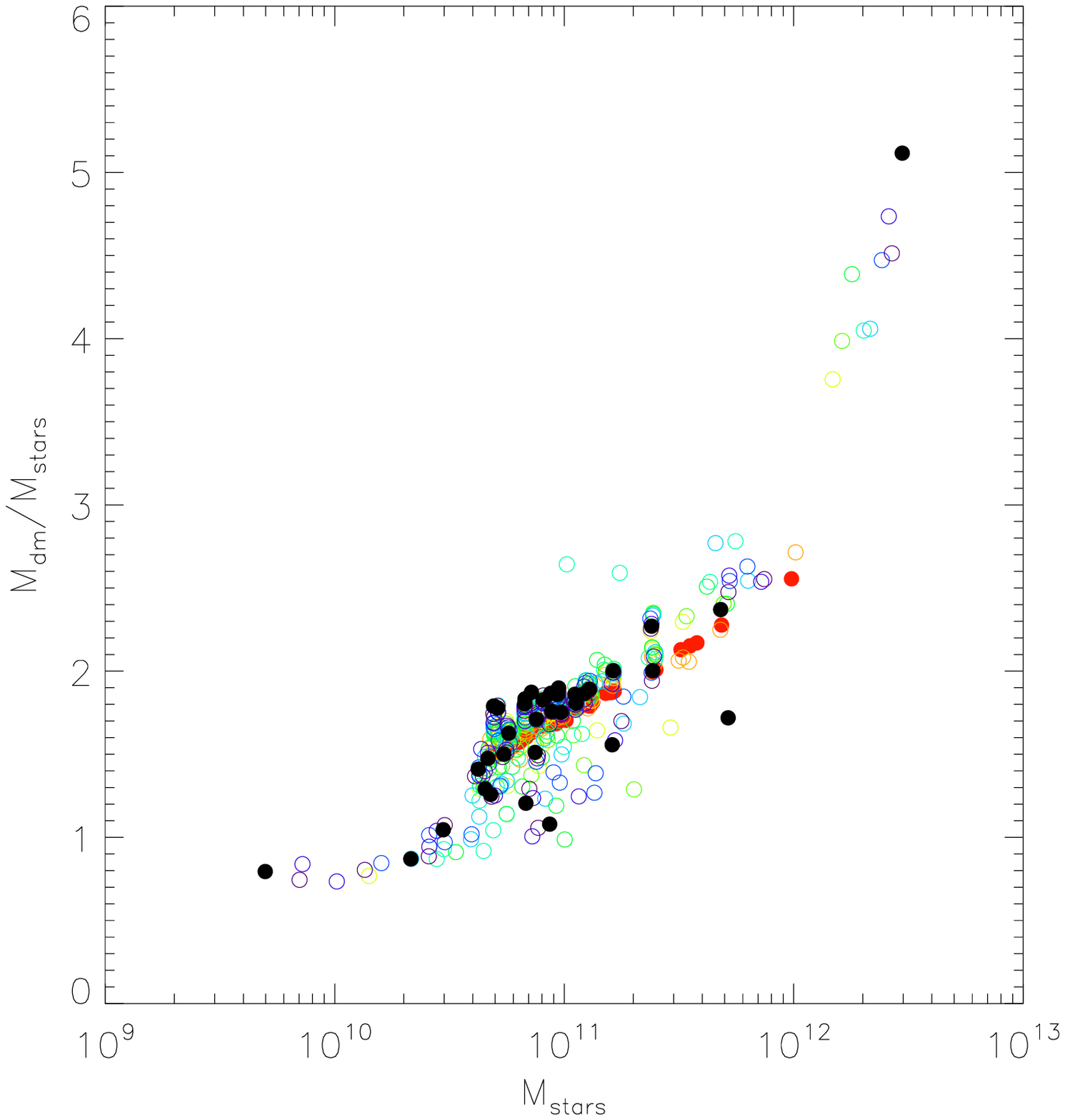}
\caption{Dark matter to stellar mass ratios vs. stellar mass. The meaning of
  the symbols is the same as in Figure 1. The left panel shows results for
  uncontracted galaxy models, while the panel on the right gives results for
  adiabatically compressed initial galaxy models.}
\end{figure*}

\label{lastpage}
\end{document}